\documentclass[rnaas]{aastex62}

\newcommand{\tess}{{\it TESS}}

\usepackage{hyperref}

\newcommand{\chicago}{Department of Astronomy and Astrophysics, University of
Chicago, 5640 S. Ellis Ave, Chicago, IL 60637, USA}
\newcommand{\sagan}{Sagan Fellow}

\begin{document}

\title{Unbiased inference of the masses of transiting planets from radial velocity followup}

\correspondingauthor{Benjamin T. Montet}
\email{bmontet@uchicago.edu}

\author[0000-0001-7516-8308]{Benjamin~T.~Montet}
\altaffiliation{\sagan}
\affiliation{\chicago}



\section{}

The \tess\ mission will detect many transiting planets 
amenable to radial velocity (RV) followup: a
\tess\ primary science requirement is the measurement of 50 masses of planets
smaller than Neptune. The transit signals that \tess\ will detect will be observed by a growing worldwide network of precision RV instruments \citep{Wright17}. Each team
will have a finite amount of observing time and will need to determine how to allocate these resources efficiently among planets.
One common strategy is to observe targets until some fractional precision on the Doppler semiamplitude is achieved. 
In practice, this means stars are observed until the planet signal is found to be nonzero at a particular significance threshold, often $6\sigma$.
Here, I show that this strategy systematically biases the inferred mass of the typical planet upward, especially when an observing campaign's 
fixed length is taken into account. This bias will have implications for studies of planetary interiors and atmospheres.

Measurements of a planet with mass $M$ are noisy, with a maximum likelihood mass estimate $\hat{M}$ and uncertainty $\sigma$.
In general,
when uncertainties are dominated by statistical measurement errors $\sigma$ 
decreases with continued observations.\footnote{This statement assumes stellar activity is modeled appropriately
and is separable from the planet signal.} When $\hat{M} > M$, a given fractional
precision will be achieved with fewer observations, so that overestimated planet masses will be considered significant before underestimated planet masses. While photon noise is equally likely to cause an inferred mass of $\hat{M} = M + \delta M$ or $\hat{M} = M - \delta M$ at a given $\sigma$, $M + \delta M$ is more likely to be ruled significant. If RV observations are then stopped once significance is
achieved, this will bias inferred masses to larger values.

To show this, I simulate a series of planetary systems\footnote{\url{https://github.com/benmontet/UnbiasedRVs}} with a signal due to one planet
on a circular orbit, as most short-period planets have low eccentricity \citep{Shabram16}. I then simulate RV observations of each star at random phase and with measurement uncertainty equal in size to the planetary signal, collecting observations until there are at least 8 RV epochs and the inferred planet mass is measured to a precision of 16\% (nonzero at $6\sigma$).

The result is shown in Figure 1. The mean inferred mass is 9\% larger than
the input masses. 59\% of all targets have overestimated masses, 24\% by more 
than 1$\sigma$ (vs. 16\% expected). 
These results assume all planets eventually have their masses measured. 
If we restrict ourselves to only the 41\% of simulated stars that had a $6\sigma$ mass measurement
after 60 epochs, 95\% of all ``publishable'' targets at the end of this simulated observing season have overestimated masses, 57\% by more than $1\sigma$
and 17\% by more than $2\sigma$. The mean inferred Doppler amplitude is 1.29 m s$^{-1}$.
The bias becomes more extreme the fewer observing epochs are permitted; the first significant detections are often the most extreme outliers.
With a systematic noise floor, this would be more extreme as detections would only be achievable when $\hat{M}$ was significantly larger than that floor, regardless of the true mass.

\begin{figure}[htp]

\centering
\includegraphics[width=.99\textwidth]{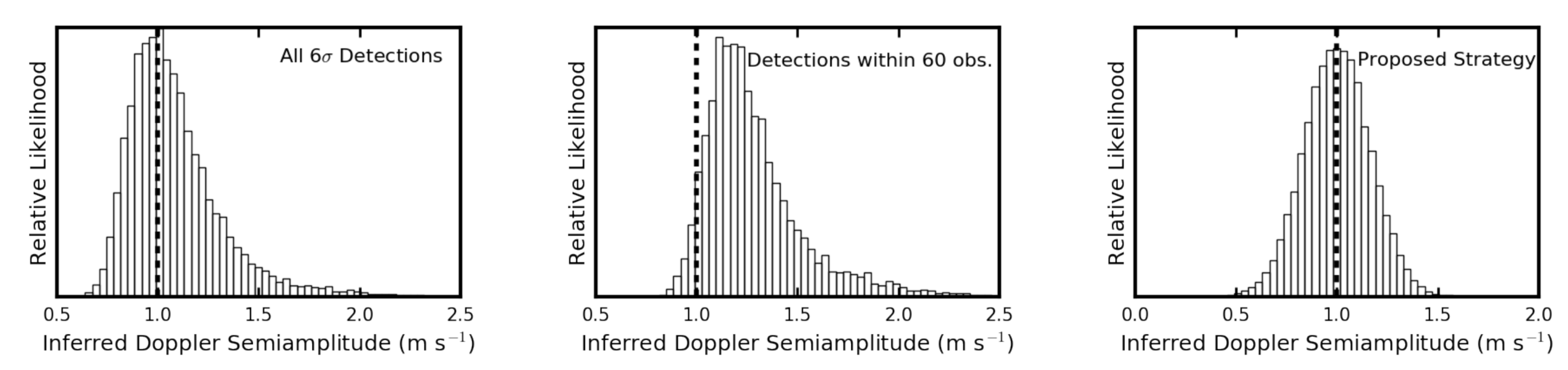}\hfill

\caption{Inferred mass of a population of planets that induce a 1 m s$^{-1}$ Doppler signal on their host star. (Left) Mass distribution for planets observed until their mass is deemed significant at the $6\sigma$ level. (Center) The same, only including planets with significant mass measurements within 60 epochs. (Right) Inferred mass distribution when a criterion to end observing does not depend on the inferred mass itself.}
\label{fig}

\end{figure}

Therefore, any observing strategy that only publishes planet masses once they reach a certain fractional precision will bias population studies of planet densities and compositions unless they also publish their non-detections \citep[e.g.][]{Marcy14} and their internal criteria for selecting when to observe---and when to stop observing---targets.

This effect can be ameliorated if, instead of observing until a certain fractional mass precision is achieved, stars are observed until a particular absolute mass precision is achieved. 
If the above simulation is carried out so that stars are observed until the Doppler semiamplitude is measured to a pre-determined absolute precision, any overestimate of $\hat{M}$ cannot affect the choice to continue observing. In this case, an inferred mass of $\hat{M} = M + \delta M$ or $\hat{M} = M - \delta M$ is still equally likely, but observing decisions will not be made from that inference. When we restrict ourselves to stars with a significant mass measurement within 60 epochs, the results are similar. Observing every follow-up target to the same
precision may be a suboptimal allocation of observing resources, but a precision could
be determined as a function of the observed planet radius as an expected RV signal can be estimated before any spectra are collected \citep[e.g.][]{Wolfgang16, Chen17}. Even if this estimate is inaccurate, it should not lead to a bias in the inferred planet mass $\hat{M}$ if the stopping criteria do not depend on $\hat{M}$ itself.

Observing teams should be encouraged to continue to observe targets after significance is achieved, and to publish their observing---and stopping---criteria for each planet and each non-detection. If stars are only observed until planet signals are detected at the 6$\sigma$ level (or any arbitrary fractional precision), the first bona fide Earth-like planet to be observed is more likely to be initially interpreted as having a composition similar to Mercury than Earth.

\acknowledgments

I thank Megan Bedell, Jacob Bean, and Emily Gilbert for providing feedback on a draft of this note, and the University of Chicago Exoplanet Journal Club for conversations that inspired this note. This work was performed under contract with the Jet Propulsion
Laboratory (JPL) funded by NASA through the Sagan Fellowship Program executed
by the NASA Exoplanet Science Institute.

\end{document}